\documentclass[preprint,12pt]{elsarticle}

\usepackage{graphicx}

\usepackage{amssymb,amsmath,color}

\biboptions{numbers,comma,square,sort&compress}

\journal{arXiv}

\begin{document}

\begin{frontmatter}



\title{Modelling of processes in nerve fibres at the interface of physiology and mathematics}


\author[]{J\"uri Engelbrecht\corref{cor1}}
\ead{je@ioc.ee}
\cortext[cor1]{Corresponding author}
\author[]{Kert Tamm}
\ead{kert@ioc.ee}
\author[]{Tanel Peets}
\ead{tanelp@ioc.ee}

\address{Department of Cybernetics, School of Science, Tallinn University of Technology, Akadeemia tee 21, 12618, Tallinn, Estonia}
\begin{abstract}
The \emph{in silico} simulations are widely used in contemporary systems biology including the analysis of nerve pulse propagation. As known from numerous experiments, the propagation of an action potential is accompanied by mechanical and thermal effects. This calls for an analysis at the interface of physics, physiology and mathematics. In this paper, the background of the model equations governing the effects in nerve fibres is analysed from a physical viewpoint and then discussed how to unite them into a system by using the coupling forces. The leading hypothesis associates the coupling to the changes of variables, not to their values or amplitudes. This hypothesis models actually the physiological mechanisms of energy transductions in a fibre. The general assumptions in modelling the processes and the properties of the coupled system of equations are described. The dimensionless mathematical model which couples the action potential with mechanical waves together with temperature effects is presented in the Appendix. This model generates an ensemble of waves including the electrical signal and mechanical and thermal effects. 
\end{abstract}


\end{frontmatter}



\section{Introduction}
\label{sec1}

Signal propagation in nervous systems is extremely important for all multicellular animals including human beings. Nerve signals control motion, behaviour and consciousness in many respects. The rich history of studies into the function of nerves is described in many overviews \cite{Bishop1956,Nelson2004}, etc and a concise history of neuroscience spiced with mathematics is given by Scott \cite{scott2002}. Contemporary studies are characterized by attempts to describe interactions between the biological and physiological functions of a nervous system across multiple scales \cite{Noble2002a,Gavaghan2006}. The better knowledge about functional behaviour of normal signalling in nerves helps to understand also the axon dysfunction, neuronal communication, synaptic efficacy, etc needed for understanding neuronal activity \cite{Debanne2011}. For this purpose, \emph{in silico} studies of biological processes are gaining more and more attention \cite{Noble2002b}. It means developing interdisciplinary approaches combining ideas of physiology, physics and mathematics in studies on nervous function.

In this paper, a possible approach is envisaged for mathematical modelling of the signal propagation in axons. In Section 2 the general background of basic model equations is analysed followed in Section 3 by physiological description of signals in nerve fibres. The assumptions made in the modelling are presented and analysed in Section 4. Further, in Section 5, a concrete mathematical model based on those assumptions is described. The discussion in Section 6 summarizes the analysis and gives the hints for further studies. The mathematical model in the dimensionless form is given in the Appendix.

\section{Mathematical modelling and basic equations}
\label{sec2}

Complexity of biological processes involving structural and functional hierarchies needs interdisciplinary approaches in order to model phenomena under consideration. This is why systems biology uses methods and tools from physics and mathematics. Sometimes such an approach is called bio-mathematical modelling \cite{Gavaghan2006}, sometimes a physical scientist’s perspective \cite{Coveney2005}, sometimes simply working at the interface of physics and biology \cite{Bialek2018}. The mathematical modelling of biological processes means casting the physiological phenomena and functional behaviour into the language of mathematics. One could recall the saying of Galileo Galilei: ”The Book of Nature is written in the language of mathematics” but using this idea is far away of being simple. Indeed, biological systems need energy exchange with the surrounding environment and represent the systems far from the thermodynamic equilibrium; they involve many transfer mechanisms that should be analyzed on the molecular level; the process may involve different time and space scales; different types of mathematical equations are needed for describing the system; different physical effects like nonlinearities, dissipation, excitability, spatiotemporal coupling, etc must be taken into account \cite{Vendelin2007}. However, mathematical modelling is not only a tool for describing biological processes but also extremely useful for performing \emph{in silico} experiments. It is said that “simple conceptual models can be used to uncover new mechanisms that experimental science has not yet encountered“ \cite{Wooley2005}. In addition, a properly chosen mathematical model has a predictive power and helps to understand the causality of the process. Without any doubt, all mathematical models need experimental validation.

The mathematical models of dynamical processes have a physical background and are derived from basic principles: conservation of mass, motion and energy together with thermodynamical conditions. Dealing with motion, the wave equation is the backbone of all the models – one of the most important equations of mathematical physics \cite{Stewart2013}. Another important equation is the diffusion equation governing the distribution of heat (or variation in temperature). Both of these classical equations of mathematical physics are well studied but too simple for describing most practical cases and need modifications in order to grasp effects which are characteristic to phenomena under consideration \cite{Vendelin2007,Engelbrecht2015}. So the mathematical models of biological processes are usually derived from those simple equations with many specific additional terms and conditions in order to better reflect realities but certainly related to basic principles. And one idea more: ``simple conceptual models can be used to gain insight, develop intuition, and understand ``how something works"" \cite{Wooley2005}. Below such a modelling is demonstrated for the case of action potential in a nerve fibre.

\section{Signals in nerve fibres, physiology and physics}
\label{sec3}

The basic component in the process of signal propagation in nerve fibres is the electrical signal called action potential (AP). Since the pioneering studies of Hodgkin and Huxley \cite{Hodgkin1945}, the electrophysiological model has been developed in many details \cite{Courtemanche1998,Clay2005,Debanne2011} etc. The process is physically complicated due to special structure of nerve fibres. Bearing in mind unmyelinated axons, a nerve fibre in mechanical terms is a tube in a certain environment. The wall of the tube has a bilayered lipid structure – biomembrane – which can be considered as a microstructured medium. Inside the tube is the axoplasmatic fluid (axoplasm), called also intracellular fluid. In mechanical terms axoplasm is a viscoelastic medium containing also cytosceletal filaments.  The biomembrane has voltage-sensitive (also mechano-sensitive) ion channels through which the ions can pass from the intracellular fluid to the environment (extracellular fluid) and vice versa. The ion concentration has an important role in emerging of an AP. The biomembrane as the cornerstone in dynamics of cells, plays an important role in biology in general, not only for nerve fibres \cite{Mueller2014}.

A signal in such a complex structure has many components and there are many experimental studies demonstrating that an AP is accompanied by mechanical effects and temperature changes \cite{Tasaki1988,Terakawa1985,Abbott1958,Howarth1968}  etc. Kaufmann \cite{Kaufmann1989} has stated that electrical action potentials are inseparable from force, displacement, temperature, entropy and other membrane variables. Indeed, the AP generates also the pressure wave (PW) in the axoplasm and a longitudinal wave (LW) in the biomembrane. The LW means a local longitudinal compression which changes the diameter of an axon (swelling) reflected by a transverse displacement (TW). This process is characterized also by heat production and temperature ($\Theta$) changes. It is a real challenge as said by Andersen et al \cite{Andersen2009}: to frame a theory that incorporates all observed phenomena in one coherent and predictive theory of nerve signal propagation. Such a theory depends on many physical properties of an axon and extracellular fluid. The AP depends on the properties of the axoplasm and ion currents, the PW depends on the properties of the axoplasm. The LW and the TW depend on the elastic properties of the biomembrane, while the $\Theta$ depends on the thermodynamical parameters of the whole system. All these quantities are coupled into a whole which reflects the complexity of the process of signal propagation in an axon.

Albert Einstein has said: “Everything must be made as simple as possible but not simpler”. In case of signal propagation in nerves, the main question is what is “simpler”? There are many physiological properties of axons \cite{Clay2005,Debanne2011}, etc which influence the process. First, all the structural elements have a certain microstructure (the lipid molecules of the biomembrane and the ion channels, the filaments and proteins in the axoplasm, etc) which play a role in signal formation and propagation. The coupling forces between the AP and the accompanying effects should reflect the physics of electro-mechanical transduction and be determined with a suitable accuracy. The thermodynamical properties of all the system should be taken into account. An overview by Drukarch et al \cite{Drukarch2018} describes the present state-of-art of analysis of this fascinating process.

Although molecular mechanisms (ion currents) influence the process, the values of main variables and axon parameters are far from the molecular range. In terms of continuum theories this means the micro- or mesoscale \cite{Gates2005} which justifies the usage of models derived from basic conservation laws of continua. Starting from the basics (see Section 2), the wave equations are the cornerstones of all the processes which propagate in media with certain finite velocities. The diffusion equations describe the processes related to heat production. Given the complicated structure of an axon, these model equations should be modified.

\section{Assumptions in modelling}
\label{sec4}

In what follows, an attempt is described following the ideas of Andersen et al \cite{Andersen2009} to build up a mathematical model which could account for main effects in signal propagation in nerve fibres. Based on classical understandings of axon physiology \cite{Clay2005,Debanne2011}, the following basic assumptions are made:\\
(i) electrical signals are the carriers of information \cite{Debanne2011} and trigger all the other processes;\\
(ii) the axoplasm in a fibre can be modelled as fluid where a pressure wave is generated due to electrical signal; here, for example, the actin filaments in the axoplasm may influence the opening of channels in the surrounding biomembrane but do not influence the generation of pressure wave in the fluid \cite{Andersen2009};\\
(iii) the biomembrane is able to deform (stretch, bending) under mechanical impact \cite{Heimburg2005};\\
(iv) the channels in biomembranes can be opened and closed under the influence of electrical signals as well as of the mechanical input; it means that tension of a membrane leads to the increase of transmembranal ion flow and the intracellular actin filaments may influence the motions at the membrane \cite{Mueller2014,Barz2013};\\
(v) there is strong experimental evidence on electrical or chemical transmittance of signals from one neuron to another \cite{Bennett2000,Hormuzdi2004} although the role of mechanical transmission is also discussed \cite{Barz2014}.

A different scheme of signal generation is proposed by Heimburg and Jackson \cite{Heimburg2005} where the mechanical wave in the biomembrane (LW) has the leading role. 

The trivial idea for constructing a mathematical model is to collect all the single equations describing the processes in a nerve fibre but these should be coupled into a joint system. The question how the coupling forces between the equations (i.e., signal components) are modelled is crucial. It is not only a mathematical construct but the forces reflect the electro-mechanical (or mechano-electrical) transduction mechanisms. 

We introduce the main hypothesis \cite{Engelbrecht2018d}: all mechanical waves in axoplasm and surrounding biomembrane together with the heat production are generated due to changes in electrical signals (AP or ion currents) that dictate the functional shape of coupling forces. The seconding hypothesis is: the changes in the pressure wave may also influence the waves in biomembrane.

It must be noted that already in the 19th century the German physiologist Emil Du Bois-Reymond has noticed that  ``the variation of current density, and not the absolute value of the current density at any given time, acts as a stimulus to a muscle or motor nerve" \cite{Hall1999}. This statement is called the Du Bois-Reymond law. 

Two essential remarks are following: (i) the changes of variables mean mathematically either the space or time derivatives; (ii) the pulse-type profiles of electrical signals mean that the derivatives have a bi-polar shape which is energetically balanced. Based on these remarks, the functional shapes of coupling forces are proposed in the form of first-order polynomials of gradients or time derivatives of variables \cite{Engelbrecht2018d,Engelbrecht2018}. 

Due to reciprocity, the mechanical effects could also influence the electrical properties of the membrane and consequently the behaviour of the AP (see Chen et al, \cite{Chen2019}). In mathematical terms it means that the model should have the variable-dependent parameters. In the present model the feedback from mechanical effects to the AP is proposed through introducing the dependence on LW into the ion current. It is possible to include some scale factors into the coupling forces but such a proposal needs more studies in order to prove a need for such an assumption. As the dynamical changes in the axon diameter (the amplitude of the TW) are many orders less than the diameter itself \cite{Tasaki1988}, the electrical properties of the fibre are taken constant during all the process.  

In addition it is assumed that the TW is derived from the LW like in mechanics of rods \cite{Porubov2003} – the transverse displacement is a space derivative of the longitudinal displacement. Concerning the posssible temperature changes, there are several possibilities to build up a mathematical model \cite{Tamm2019}.

It must be stressed that the assumptions described above are logically consistent within the present framework of axon electrophysiology \cite{Clay2005,Debanne2011} and experimental studies on accompanying effects. What is important – all the components of a wave ensemble including the AP are coupled. The model serves the contemporary trends in computational biology and could help experimentalists to check the possible transduction mechanism \cite{Wooley2005}.

\section{Model equations}
\label{sec5}
The mathematical analysis of the AP is usually based on the celebrated Hodgkin-Huxley (HH) equation which is derived from a telegraph equation neglecting inductivity \cite{Hodgkin1945}. The wave character of this benchmark model is preserved by adding the influence of ion current, so the result is a reaction-diffusion type equation. Three phenomenological variables govern the “turning on” and “turning off” the membrane conductance and are described by relaxation equations. The values of all the parameters needed for calculations are determined by ingenious experiments. As a result the profile of an AP has a typical asymmetric shape with an overshoot responsible for the relaxation length between pulses. An AP can be generated by an input over a certain threshold value. Contemporary studies \cite{Clay2005,Debanne2011} have demonstrated that this model, complicated as it is, should be made even more complicated for describing effects like neuronal timing and synaptic efficacy, etc. However, in order to describe a correct functional shape of an AP, a simpler model was proposed by Nagumo et al \cite{Nagumo1962} which has taken only one ion current into account. The main properties of the AP described above, are preserved and that is why this model called after FitzHugh-Nagumo (FHN) is used in the robust model proved by many calculations like for example those by Engelbrecht et al \cite{Engelbrecht2018d,Engelbrecht2018}. 

The second important component in the signal complex is the LW in the biomembrane. Due to the complicated internal structure (microstructure) of a biomembrane made of lipids, a simple wave equation is not sufficient for describing the wave motion but nonlinearity and dispersion must be also taken into account. Heimburg and Jackson \cite{Heimburg2005} have derived a wave equation (HJ equation for short) where the compressibility of the lipid molecules has an impact on velocity and an ad hoc dispersive term describes dispersion. Engelbrecht et al \cite{EngelbrechtTammPeets2014} have shown that based on mechanics of microstructured media dispersive effects should be described by two terms – one like proposed by Heimburg and Jackson \cite{Heimburg2005} reflecting elasticity and another term, reflecting inertia of lipid molecules. The governing equation for the LW in biomembranes is then described by the improved Heimburg-Jackson (iHJ) equation. The possible solutions of this Boussinesq-type iHJ equation are analysed by Engelbrecht et al \cite{Engelbrecht2017}. The remarkable property of the HJ and iHJ equations is that they possess solitonic solutions \cite{Heimburg2005,EngelbrechtTammPeets2014}. However, the influence of the inertia of lipids makes the soliton narrower. It must be stressed that the LW is generated by an AP and the formation of a soliton from such a special input takes time and depends on the energy of the input. 
 
The third component in the signal complex is the PW in the axoplasm. As far as axoplasm can be treated as a viscoelastic fluid, the classical Navier-Stokes equations can be used. However, given the very small amplitudes of the PW \cite{Tasaki1988,Terakawa1985}, nonlinear effects can be neglected and that is why the classical wave equation with a viscous term added can effectively be used.

These three components in the signal complex are all wave-like and possess definite finite velocities. In such a way these waves are the primary components in the ensemble of waves which are coupled into a whole by coupling forces modelled following the main hypothesis (see Section 4).

The remaining components (the TW and temperature $\Theta$) have no finite velocities of themselves and are derived from primary ones. This is the reason why these components in an ensemble are secondary \cite{Engelbrecht2018arXiv}.

The TW is derived from the LW by derivation following the theory of rods \cite{Porubov2003}. The derivation of a suitable model for temperature $\Theta$ generation is more complicated because there is no widely accepted consensus about the mechanism of the heat generation. The simulations show \cite{Tamm2019} that there are several mathematical models that could match various experimental results. It seems, however, that a diffusion equation with a source term based on the conservation of energy, is best suited for modelling the temperature change accompanying the AP. The source term like in the case of primary components, depends on changes of other variables. 

The collection of all equations needed for describing an ensemble of waves in an axon involving primary and secondary components is presented in the Appendix.

\section{Discussion}
\label{sec6}
What is described above, is a straightforward modelling of the complex process in nerve fibres. Without any doubt, this is a robust model and serves first of all as a proof of concept. However, after demonstrating that the basic equations of mathematical physics in their modified forms can be used to build up the backbone of nerve pulse dynamics, the further studies could enlarge significantly the description of physiological effects. Here we have been working on the integrative level \cite{Noble2002a} trying to unite the different functional elements into a whole.
 
The main points in the proposed model can be listed as follows:\\
(i) the proposed model is a robust one but describes qualitatively correctly the profiles of essential signal components in a nerve fibre; as a matter of fact, an ensemble of waves propagates in the fibre;\\
(ii) the AP is described by the FHN model which takes only one ion current into account which is assumed to be similar to the sodium current \cite{Nagumo1962}; a possibility to distinguish between the voltage-sensitive and mechano-sensitive ion channels \cite{Mueller2014} is accounted by different parameters governing the ion current;\\
(iii) the model at this step can be taken as a proof of concept and the possible next step in modifying the analysis is to use the HH model instead of the FHN model;\\
(iv) the important novel hypothesis for constructing the joint model is related to the coupling forces between the single components of the full signal;\\
(v) it is assumed that the coupling forces depend on changes of coupled signals (pulses) i.e., on their derivatives, not on their amplitudes;\\
(vi) by generating the AP from an initial input, the other waves are generated due to the coupling forces;\\
(vii) the properties of the TW which has been measured in several experiments depend on the mechanical properties of the biomembrane through the LW;\\
(ix) there are several possible ways to model the temperature changes accompanying the AP demonstrated in experiments.

It seems that the crucial problem in all mathematical models is related to the transduction of energy from one component in the ensemble to another. Several possibilities to model the coupling have been proposed. El Hady and Machta \cite{ElHady2015} have elaborated a mechanism of electro-mechanical coupling based on the assumption that the potential energy of the process is stored in the biomembrane and kinetic energy in the axoplasmic fluid resulting in mechanical surface waves. The HH model is used for describing the AP and the force exerted on the biomembrane is taken proportional to the square of the voltage while the axoplasmic fluid is described by the linearized Navier-Stokes equation. Another coupled model of electrical and mechanical signals based on spring-dampers (dashpots) system is proposed by Jerusalem et al \cite{Jerusalem2014}. Johnson and Winslow \cite{Johnson2018} have discussed on the basis of physiological effects, how to unite the HH model on the macroscopic level and the soliton model \cite{Heimburg2005} on the microscopic level. In their approach the Na$^{+}$  ion current is responsible electro-mechanical interaction. Chen et al \cite{Chen2019} have modelled the coupling of electrical and mechanical (transverse displacement) effects by using the idea of flexoelectricity. They have used the HH model for the AP and the linear viscoelastic Maxwell-type model for the biomembrane where due to the flexoelectric effect the AP generates a strain gradient bending the membrane upwards. This means that the AP is accompanied by a transverse displacement. The flexoelectric force depends on the change of the membrane potential. In this analysis the biomembrane is a homogeneous tube with elastic and viscous properties. The numerical calculations by means of the finite element method permit to solve direct (a mechanical effect generated by an AP) and reverse (an AP generated by the mechanical effect) problems. This means that reciprocity of processes is possible. The processes in both myelinated and unmyelinated axons were analyzed.

As stressed by Coveney and Fowler \cite{Coveney2005}, coupled models across several processes could provide also the route for calculating many unknown parameters. That explains why systems biology turns so much attention to the modelling of biological complexity involving molecular and continuum approaches. The \emph{in silico} experiments permit to cover a large area of possible physical parameters in order to find suitable sets verified by experiments \emph{in vivo} or \emph{in vitro}. This is also the case of nerve signals where the molecular structure of the biomembrane affects the process and the conjectured coupling forces need further studies and quantization. Although much is known about the fascinating process of nerve signalling, the full picture needs further concerted efforts of experimentalists and theoreticians. A recent  detailed overview \cite{Drukarch2018} reflects the contemporary insights in this field that could lead to “the formulation of a more extensive and complete conception of the nerve impulse”.

In modelling described above, the path was from structures to dynamics trying to grasp the leading effects in mathematical terms. Such an approach is certainly robust which is widely used in systems biology \cite{Kitano2002}. The further modifications of the model should turn to details like accounting of the various ionic currents, membrane channels fluctuations, general oxygen consumption, CO$_{2}$ output, the influence of carbohydrates and membrane proteins, etc together with elaborating the mechanisms of heat production. This will enhance the predictive power of the model which at this stage is related to qualitative estimations. These estimations correspond to known experimental results but need quantitative analysis.

The modelling at the interface of physiology and mathematics casts more light to the fascinating phenomenon of nerve pulse propagation. The structure of mathematical equations, especially of coupling forces helps to understand the causality of effects. Typically to biological processes, the coupling of various phenomena or fields must be taken into account. In this case one could modify the Du Bois-Reymond law: in dynamical processes every variation of fields acts as a stimulus to other fields. As explained above, mathematically the variation means the space or time derivatives of variables.  Last but not least, as stressed by Konrad Kaufmann, one cannot forget physical background of biological processes. That is why for such a complex biological process as the propagation of signals in nerves the modelling associated with basic equations of mathematical physics is justified.

\section*{Acknowledgements}
This research was supported by the Estonian Research Council (project IUT 33-24). 


\section*{The mathematical model}
\label{appendix}

The mathematical model in the dimensionless setting presented below has been elaborated and used for numerical simulations in earlier publications \cite{EngelbrechtTammPeets2014,Engelbrecht2017,Engelbrecht2018d,Engelbrecht2018,Engelbrecht2018arXiv,Tamm2019}. Here and further subindex $X$ denotes partial derivative with respect to space and $T$ denotes partial derivative with respect to time. 
The primary (wave-like) components:\\
\textbf{AP} (the FHN equation):
\begin{equation}
\label{FHNeq}
\begin{split}
Z_{T} &=  D Z_{XX} - J + Z \left( Z -  G_1  - Z^2 +  G_1  Z \right); \\
J_{T} &=  \varepsilon \left(  G_2  Z - J \right),
\end{split}
\end{equation}
where $G_i = a_i + b_i$ and $b_i = -\beta_i U$. Here
$Z$ -- action potential, $J$ -- ion current, $\varepsilon$ -- the time-scales difference parameter, $a_i$ -- ``electrical'' activation coefficient, $b_i$ -- ``mechanical'' activation coefficient and $U$ -- longitudinal density change from lipid bi-layer density model \eqref{iHJeq} and $\beta_i$ is an coupling coefficient.\\
\textbf{PW} (the wave equation with viscous and source terms):
\begin{equation}
\label{Peq}
P_{TT} = c_{f}^{2} P_{XX}  - \mu P_T + F_1(Z,J,U),
\end{equation}
where $P$ -- pressure, $\mu$ -- viscosity coefficient. The $F_1$ is the coupling term accounting for the possible influence from the AP and TW.\\
\textbf{LW} (the iHJ equation with a source term):
\begin{equation}
\label{iHJeq}
\begin{split}
U_{TT} &=  c^2 U_{XX} + N U U_{XX} + M U^2 U_{XX} +
   N U_{X}^{2} + \\ &+ 2 M U U_{X}^{2} -
   H_1 U_{XXXX} + H_2 U_{XXTT} + F_2(Z,J,P),
\end{split}
\end{equation}
where $U = \Delta \rho$ is the longitudinal density change, $c$ is the sound velocity in the unperturbed state, $N,M$ are nonlinear coefficients, $H_i$ are dispersion coefficients. Where $H_1$ accounts for the elastic properties of the bi-layer and $H_2$ the inertial properties. The $F_2$ is the coupling term accounting for the possible influence from the AP and PW.

The secondary components:\\
\textbf{TW} (related to the LW):
\begin{equation}
W \propto U_X,
\end{equation}
where $W$ -- transverse displacement, $U$ -- the LW.
\\
$\mathbf{\Theta}$ (the diffusion equation with a source term):
\begin{equation}
\label{HHeq}
\Theta_{T} = \alpha \Theta_{XX} + F_3(Z,J,U),
\end{equation}
where $\Theta$ -- temperature, $\alpha$ -- thermal conductivity coefficient. The $F_3$ is the coupling term accounting for the possible influence from the AP and LW.
\\
The initial condition is in terms of Z:
\begin{equation}
\begin{split}
Z(X,0) &= A_{z} \mathrm{sech}^2 B_{o} X; \quad Z_T(X,0) = 0;\\ 
Z(X,T) &= Z (X + 2 L m \pi,T); \quad  m = 1,2,\ldots ,
\end{split}
\end{equation}
while other variables have initially zero values and the same periodic boundary conditions. Here $A_{z}$ is the initial AP amplitude, $B_{o}$ is the pulse width and $L$ is the number of $2\pi$ sections in space.






%

\end{document}